\documentstyle[12pt]{article}

\def\bq{\begin{eqnarray}}
\def\eq{\end{eqnarray}}

\begin{document}
\title{\bf One loop fermion contribution to 
the effective potential for constant 
SU(2) Yang-Mills fields}
\author{NISTOR NICOLAEVICI\\
\it Technical University of Timi\c soara, Department of Physics,\\
\it P-\c ta Hora\c tiu 1, RO-1900 Timi\c soara, Romania}
\maketitle

\begin{abstract}
We obtain a series expansion for the one loop fermion contribution to the 
effective potential for constant $A^a_\mu$ fields in the SU(2) theory with a 
massive fermionic doublet. The series converges for bounded electric fields 
in terms of the magnetic fields and the gauge potentials. One finds that 
spontaneous fermion pair creation may be absent for arbitrary strong pure 
electric fields, with an appropriate choice of the classical currents.

\end{abstract}


\newpage
Fifty years ago, in a famous Q.E.D. calculation \cite{schwinger} J. Schwinger 
obtained 
the $\hbar$ contribution to the electromagnetic effective Lagrangian for 
constant (classical) fields. 
It seems there are no similar results for fermionic contributions in 
non-abelian 
theories. Our intention here 
is to briefly present a calculation for the analogous quantity in the SU(2) 
theory with a $j=\frac{1}{2}$ massive fermionic multiplet. We restrict 
to this representation for mainly technical reasons: 
for this representation calculations significantly simplify due to 
the properties of Pauli matrices (see the next footnote). We mention from 
start, that unlike in Ref. \cite{schwinger}, our result is not of maximum 
generality by failing to apply to certain field configurations, as it will 
become apparent from our computational procedure.

Two aspects specific to non-abelianity are of direct relevance 
in what follows. 
First, constant gauge potentials do not necessarily imply 
the vanishing of the field strengths. We shall do the calculations 
precisely for this case, $i.e$. (all notations are conventional) 
\bq
A^a_\mu(x)=const,
\label{co}
\eq
so that we shall actually obtain the $\hbar$ effective potential. 
Second, constant field strengths do not entail the vanishing of the 
corresponding classical currents. (In fact, under conditions (\ref{co}) 
a null current necesarilly means a pure gauge.) 
Also in contrast to the abelian case, the classical equations 
determine the currents in terms of the field strengths $and$ the potentials 
(via the covariant derivatives). Physically, this is expected to manifest 
in the explicit dependence of the quantum contribution on both of these 
quantities, as we shall indeed find out.

We begin by writing the basic Lagrangian as
\bq
{\cal L}=-\frac{1}{4}F^a_{\mu\nu} F^{a\mu \nu}+
i\bar\psi \gamma^{\mu}
(\partial_\mu-iA^a_\mu\frac{ \sigma_a}{2})
\psi-m\bar\psi\psi,
\label{lag}
\\
F^a_{\mu \nu}=\partial_\mu A_{\nu}^a-
\partial_\nu A_{\mu}^a+\varepsilon_{abc}
A^b_\mu\,A^c_\nu,\quad a,b,c=1,2,3. 
\eq
We included for simplicity 
the coupling constant 
into the definitions of $A^a_\mu$, $F^a_{\mu\nu}$. As a first step, we need 
the one loop effective action for the classical $A^a_\mu$ field. 
This is most directly obtained in the path integral formalism by 
integrating out the $\psi$ field 
\bq
S_{eff}=-i\ln\int {\cal D} \bar \psi {\cal D}\psi 
\exp i S[A^a_\mu, \bar\psi, \psi],
\label{psiout}
\eq
($S$ is the classical action). It is clear that ghost terms are 
unnecessary here. The integral is a Gaussian in Grassmannian 
variables, which can be evaluated by standard methods \cite{rivers}. 
One finds for the $\hbar$ fermionic contribution
\bq
S_{eff}^{1,f}=-i\ln \mbox{Det}\,
((\gamma^\mu(P_\mu-A_\mu^a \frac{\sigma_a}{2})-m),
\label{s1}
\eq
where Det is the determinant in the functional sense and $P_\mu$ are the
translation operators in coordinates space. $m$ tacitly incorporates the
$i\epsilon$ prescription. Our subsequent efforts will be to extract 
information from eq. (\ref{s1}).

Using the charge conjugation matrix $C\gamma_\mu C^{-1}=-\gamma_\mu^T$ and the 
properties of the determinant, one can eliminate the gamma matrices 
(see Ref. \cite{itzykson}) to obtain
\bq
S_{eff}
^{1,f}
=-\frac{i}{2}\ln \mbox{Det}\,
((P_\mu-A^a_\mu\frac {\sigma_a}{2})^2-m^2).
\label{s2}
\eq
We use next the ln Det=Tr Ln formula and the Schwinger representation for the 
logarithm. When tracing over space-time coordinates, translational invariance 
allows separation of the four-volume factor $\int d^4x$. 
The trace over spinorial indices yields simply a 4 factor. Tracing over the 
group indices amounts to evaluate, after expanding the square in 
eq. (\ref{s2}),
the trace of a SU(2) matrix\footnote{For $j>1/2$ representations, 
anticommutations of the group generators will generally lead to SU(2$j$+1) 
elements. The quantity under square root in eq. (\ref{integral}) becomes then 
a generalized quadratic form in $p_\mu$, making the $d^4p$ integral more 
difficult.} in 
exponential parameterization. One finally arrives to the following 
expression for the effective potential ($A^2=A^a_\mu A^{a \mu}$):
\bq
V_{eff}^{1,f}=
-4i\int_0^\infty\frac{ds}{s}\exp is(-m^2+A^2/4)
\nonumber\\
\times\int \frac{d^4p}{(2\pi)^4}
\cos s\sqrt{(p^\mu A^a_\mu)(p^\nu A_\nu^a)}
\exp isp^2-\mbox{ST.}
\label{integral}
\eq
ST stands for a subtraction term corresponding to the expression 
evaluated at null potentials. 
Consider a four vector $K_\mu$ with $K^\mu K_\mu=A^2$. One can 
easily verify that ST can be obtained by replacing the cos factor in the second 
integral with 
\bq
\cos s(p^\mu K_\mu)\label{cos}
\eq
and leaving $A^2$ unchanged. We shall consider this quantity to perform the 
subtraction, instead of that resulting by simply setting $A^a_\mu=0$. This 
trick will make apparent at the end of the calculations that, 
as expected, the potential vanishes for $F^a_{\mu\nu}=0$. 

We rotate next the $p_\mu$ integration contours as 
(latin indices $i,j,k$ run over $1,2,3$) 
\bq
p^0\rightarrow e^{i\frac{\pi}{4}}p^0, \quad
p^i\rightarrow e^{-i\frac{\pi}{4}}p^i,
\eq
making the exp factor in eq. (\ref{integral}) a Gaussian in 
$p_0$, $p_i$, and shift to four-dimensional spherical coordinates. At this 
point, unfortunately, a further closed evaluation becomes rather 
impossible: with a series of integration by parts the radial integration 
can be written $\int_0^\infty dz \sin\,az \exp -sz^2$ 
($a$ contains the angular and $A^a_\mu$ dependence) 
letting itself expressed \cite{grad} in terms of Erfi$(a/2\sqrt{s})$, which 
makes the angular integration untractable. We shall continue by expanding the 
cosines in Taylor series, 
with the obvious mention that by doing so we lost maximum generality. 
The validity of the 
result will be discussed a little further.

After performing the radial integration, each term yields a finite result for 
$s>0$ fixed. To
account for the angular integration, it is convenient to define the matrices
\bq
\omega_{\mu\nu}^A=\tilde A^a_\mu \tilde A^a_\nu,
\quad
\omega_{\mu\nu}^K=\tilde K_\mu \tilde K_\nu,
\eq
where
\bq
\tilde A^a_0=e^{i\frac{\pi}{4}} A^a_0,\quad \tilde A^a_i=e^{-i\frac{\pi}{4}}
A^a_i,\quad \mbox{similarly for } \tilde K_\mu,
\eq
along with the quantities
\bq
V_n=i^n \frac{n+1}{2\pi}
\left\{
\int d\Omega_n
(\omega^A_{\mu \nu} n_\mu n_\nu)^n
-\int d\Omega_n
(\omega^K_{\mu \nu} n_\mu n_\nu)^n
\right\}.
\label{omega}
\eq
Notations in eq. (\ref{omega}) are as follows: $n_\mu$ is the unit four-vector
$n_\mu n_\mu=1$, the lowered indices indicating summation with respect
to the Euclidean metric. The integrations extends over the Euclidean 3-sphere, 
with $d \Omega_n$ the infinitesimal angle around $n_\mu$.

The behavior with respect to $s$ integration for 
$s\rightarrow 0$ formally 
splits $V_{eff}^{1,f}$ into an infinite and a 
finite part. The infinite part reads 
\bq
V^{1,f}_{inf}=-\frac{1}{4\pi^2}
\sum_{n=0}^2 
i^n
\frac{n!}{(2n)!}\,
V_n
\int_0^\infty \frac{ds}{s^{3-n}}
\exp is(-m^2+A^2/4).
\eq
The first two terms, however, vanish due to the angular integrations 
(cf. eq. (\ref{radicals}) below) 
\bq
V_0=V_1=0.
\eq
For $n=2$, the $s$ integral diverges logarithmically, but this is also no 
trouble since 
\bq
V_2=-\frac{1}{4}F^a_{\mu\nu}F^{a\mu \nu}.
\eq
The divergence is proportional to the pure Yang-Mills Lagrangian,
which amounts to a wave function and coupling constant renormalization. 
Discarding an infinite contribution and imposing the quantum contribution to 
vanish for $A^a_\mu\rightarrow 0$, one obtains the renormalized quantity
\bq
V^{1,f}_{ren}=\frac{1}{12\pi^2}\ln \left(1-\frac{A^2}{4m^2}\right)
F^a_{\mu\nu}F^{a\mu \nu}.
\eq

For $n\geq 3$, terms are individually finite and give
\bq
V^{1,f}_{fin}=\frac{m^4}{4\pi^2}
\left(1-\frac{A^2}{4m^2}\right)^2
\sum_{n=3}^\infty
\frac{(n-3)!\, n!}{(2n)!}\,\hat V_n,
\label{series}
\eq
where we introduced the adimensional quantities 
\bq
\hat V_n=
\frac{V_n}{\left(m^2-A^2/4\right)^n}.
\eq
$\hat V_n$ for $n$ arbitrary can be obtained by evaluating 
the Gaussians 
\bq
\int d^4 x\, \exp [x_\mu x_\nu (i\sigma \hat\omega^{A,K}_{\mu \nu}-
\delta_{\mu\nu})],
\quad
\hat \omega^{A,K}_{\mu \nu}=
\frac{\omega^{A,K}_{\mu \nu}}{m^2-A^2/4},
\label{techne}
\eq
and taking the $n$th $\sigma$ derivative at $\sigma=0$. The
result is 
\bq
\hat V_n=\frac {1}{n!}
\left (\frac{\partial}{\partial \sigma} \right)^n
\left ( \frac{1}{\sqrt {1+Z_1\sigma+Z_2\sigma^2+Z_3\sigma^3}}-
\frac{1}{\sqrt {1+Z_1\sigma}}
\right ) 
\biggl \vert_{\sigma=0},
\label{radicals}
\eq
where $Z_1$, $Z_2$, $Z_3$ are the (globally) gauge invariant quantities
\bq
Z_1=\frac{A^2}{m^2-A^2/4},\\
\nonumber\\
Z_2=\frac{1}{2}\frac{F^a_{\mu\nu}F^{a\mu \nu}}{\left(m^2-A^2/4\right
)^2},
\label{blast}\\
\nonumber\\
Z_3=\frac{1}{6}\frac{
\varepsilon_{abc}F^{a\nu}_\mu F^{b\lambda}_\nu F^{c \mu}_\lambda}
{\left(m^2-A^2/4\right)^3}.
\label{last}
\eq
The polynomials under the radicals are, in order, the determinants 
of the matrices in the exponent in eq. (\ref{techne}) corresponding to 
$A$, $K$ superscripts respectively. This concludes our calculation. Note that 
the indeterminacy introduced by $K_\mu$ disappeared.

It is transparent from eqs. (\ref{radicals})-(\ref{last}) that for 
vanishing field strengths all $\hat V_n$ are zero, 
which makes $V^{1,f}_{eff}=0$. One sees also that, as already pointed out, an 
explicit dependence on the potentials survives through $A^2$.

We come now to the question of the convergence of 
series (\ref{series}). Numerical factors for $n\rightarrow \infty$ imply 
it behaves as 
\bq
\sum^\infty_{n} 4^{-n}\hat V_n.
\label{pure}
\eq
We shall demand, for safety, convergence of each sum defined by 
the two square roots in eq. (\ref{radicals}). For the second 
one, one readily recognizes the $Z_1$ power expansion of $\sqrt{1+Z_1/4}$, 
so we have to set
\bq
 \vert Z_1 \vert <4,
\eq
or
\bq
-\infty < A^2 <2m^2.\label{abound}
\eq
For the first sum, we proceed as follows. One sees that $\lambda_0=0$ 
is always an eigenvalue\footnote {There always exists a non-zero vector 
$\tilde X_ \mu$ such that $\tilde A^a_\mu \tilde X_\mu=0$ for all $a$, hence 
$\hat\omega^A_{\mu \nu} \tilde X_\nu$=0.} of $\hat \omega^A_{\mu \nu}$, and 
let us denote $\lambda_1$, $\lambda_2$, $\lambda_3$ the remaining 
ones. This makes $-i\lambda_{1,2,3}^{-1}$ the roots of the cubic polynomial 
in eq. (\ref{radicals}). Then the sum in eq. (\ref{pure}) can be organized as
\bq
\sum_{n_1,n_2,n_3}^\infty \prod_{r=1}^3
\frac {1}{4^{n_r}\, n_r!}
\left (\frac{\partial}{\partial \sigma} \right)^{n_r}
\!\!\!\frac{1}{\sqrt {1-i\lambda_r \sigma}}\,
\biggl \vert_{\sigma=0},
\eq
and convergence similarly requests
\bq
\vert \lambda_r \vert<4, \quad r=1,2,3.\label{cond}
\eq

Let us introduce electric and magnetic fields
\bq
E^a_i=F_{0i}^a,\quad B^a_i=\frac{1}{2}\epsilon_{ijk} F_{jk}^a.
\eq
A closer inspection of inequalities (\ref{cond}) reveals that for $A^2$ 
and ${\bf B}^a$ fixed, they define a $bounded$ set in the 
${\bf  E}^a$ space. Analysis in general case proves 
fairly involved, as solutions of a cubic equation have to be 
considered explicitly. We shall 
restrict to the special 
situation\footnote{An immediate example are the pure electric fields 
${\bf  B}^a=0$.} $Z_3=0$, or
\bq
{\bf E}^1\cdot ({\bf  B}^2\times {\bf B}^3)
+{\bf E}^2\cdot ({\bf  B}^3\times {\bf B}^1)
+{\bf E}^3\cdot ({\bf  B}^1\times {\bf B}^2)
-{\bf B}^1\cdot ({\bf  B}^2\times {\bf B}^3)=0,\nonumber\\
\eq
for which things become particularly simple. One finds that 
conditions (\ref{cond}) reduce, for all $r$, to the single 
inequality
\bq
E^2<B^2+16 m^4
\left (1-\frac{A^2}{4m^2} \right)
\left(1- \theta(A^2)\, \frac{A^2}{2m^2}\right),\label{ebound}
\eq
where
\bq
E^2={\bf  E}^a\cdot {\bf  E}^a,\quad B^2={\bf  B}^a\cdot {\bf  B}^a,
\eq
and $\theta$ is the Heaviside step function. We mention that in arriving to 
eq. (\ref{ebound}) we used, in addition, the inequality 
\bq
E^2\geq B^2-(A^2)^2/3, 
\eq
which results as a direct consequence of definitions of 
${\bf E}^a$, ${\bf B}^a$ in terms of $A^a_\mu$.

Relation (\ref{ebound}) 
deserves a comment: it implies that for $B^2$ fixed 
and $A^2<0$ variable, there is generally $no$ definite upper bound for 
$E^2$. With a proper choice of the gauge potentials (or of the classical 
currents), one can set it arbitrarily high. 
One can easily check, for example, that for pure electric fields $A^2$ can be 
taken arbitrarily large and negative, while keeping ${\bf  E}^a$ constants.

It is interesting to note that convergence is always assured for pure magnetic 
fields ${\bf E}^a=0$, no special conditions assumed. One can always consider 
in this case $A^a_0=0$, automatically satisfying eq. (\ref{abound}). (Non-zero 
temporal components demand ${\bf B}^a=0$ too. One can directly show then the 
second 
integral in eq. (\ref{integral}) is independent of $A^a_\mu$.) One further 
notes that $i\lambda_{1,2,3}$ are real and 
positive, as the matrix $i\hat\omega^A_{jk}$ is real, symmetric and 
positive definite. From the positivity of trace one has 
\bq
0<\sum_{r=1}^3 i\lambda_r=-Z_1<4,
\eq
which means that inequalities (\ref{cond}) are also assured.

We end with the obvious observation that 
$V_{eff}^{1,f}$ contains no absorptive part. It follows that for field 
configurations to which 
our result applies, no spontaneous fermion pair creation 
occurs. Convergence conditions formulated 
above can be regarded as sufficient conditions to forbid this phenomenon, 
to order $\hbar$. In particular, one comes to the non-trivial 
conclusion that fermion pair creation does not necessarily take place for 
pure electric fields, in contrast to the well known case of quantum 
electrodynamics. Absence of the absorptive part in our result is, in fact, 
not surprising: one knows that particle creation in time independent fields 
is essentially a non-perturbative effect \cite{mamaev}. In our calculation, 
this quality was clearly lost (in a strict sense, at least) with the Taylor 
expansion of the cos factors in eq. (\ref{integral}).

\bigskip
\noindent
\section*{Acknowledgments}
I thank to Atilla Farkas for reading the manuscript.

\end{document}